\documentclass[doublecol]{epl2}
\usepackage[utf8]{inputenc}
\usepackage{amsmath,amssymb,graphicx,color,url}
\newcommand{\T}{^{\mathrm{T}}}
\newcommand{\avg}[1]{\langle #1\rangle}
\newcommand{\vek}[1]{\boldsymbol{#1}}
\newcommand{\norm}[1]{\left\|#1\right\|_2}
\title{Ranking users, papers and authors in online scientific communities}
\author{H. Liao$^1$, R. Xiao$^1$, G. Cimini$^{1,2}$, M. Medo$^1$}
\shortauthor{H. Liao \etal}
\institute{
  \inst{1} Physics Department, Chemin du Musée 3, University of Fribourg, 1700 Fribourg, Switzerland\\
  \inst{2} Grupo Interdisciplinar de Sistemas Complejos (GISC), Departamento de Matemáticas, Universidad Carlos III de Madrid, 28911 Leganés, Madrid, Spain}
\abstract{The ever-increasing quantity and complexity of scientific production have made it difficult for researchers to keep track of advances in their own fields. This, together with growing popularity of online scientific communities, calls for the development of effective information filtering tools. We propose here a method to simultaneously compute reputation of users and quality of scientific artifacts in an online scientific community. Evaluation on artificially-generated data and real data from the Econophysics Forum is used to determine the method's best-performing variants. We show that when the method is extended by considering author credit, its performance improves on multiple levels. In particular, top papers have higher citation count and top authors have higher $h$-index than top papers and top authors chosen by other algorithms.}
\begin{document}
\maketitle

\section{Introduction}
Science is not a monolithic movement, but rather a complex enterprise divided in a multitude of fields and subfields, many of which enjoy rapidly increasing levels of activity~\cite{Radicchi08,Lari08}. Even sub-disciplines have grown so broad that individual researchers cannot follow all possibly relevant developments. Despite swift growth of online scientific communities (such as ResearchGate, Mendeley, Academia.edu, VIVO, and SciLink)~\cite{Maxmen10} which facilitate social contacts and exchange of information, finding relevant papers and authors still remains a daunting task, especially in lively research fields.

More generally, reliance of the modern society on computer-mediated transactions has provoked extensive research of reputation systems which compute reputation scores for individual entities and thus reduce the information asymmetry between the involved parties~\cite{Resnick00,Sabater05}. Perhaps more important than the immediately useful information is the proverbial shadow of the future---incentives for good behavior and penalties against offenses---generated by these systems~\cite{Axelrod,Massum04}. Reputation systems are now an organic part of most e-commerce web sites~\cite{Josang07} and question \& answer sites~\cite{Hanrahan12}. Complex networks~\cite{Newman_book} have provided a fruitful ground for research of reputation systems with PageRank~\cite{PR,Franc11} and HITS~\cite{HITS} being the classical examples. In~\cite{Fujimura05}, the authors extended HITS by introducing authority score of content providers and apply the resulting EigenRumor algorithm to rank blogs. Building on a bipartite version of HITS~\cite{coHITS}, \cite{Hao12} presents a QTR algorithm suited for online communities. This algorithm co-determines item quality and user reputation from a multilayer network which consists of a bipartite user-item network and a monopartite social network.

We propose here a reputation algorithm designed especially for online scientific communities where researchers share relevant papers. We first simplify the QTR algorithm by neglecting the social network among users. This simplification reflects the fact that trust relationships are often not available and allows us to better study the algorithm's output with respect to the remaining parameters. We then extend the algorithm by introducing author credit which is however computed differently than in the previously-mentioned EigenRumor. Since author credit is co-determined from the same data as paper quality and user reputation, this extension preserves an important advantage of QTR: Its reliance on implicit ratings which are easier to elicit than explicit ratings~\cite{Josang07}. We use various standard metrics of research productivity (citation count, impact factor, and $h$-index) to demonstrate that the new algorithm outperforms other state-of-the-art algorithms.

\section{Algorithms}
An online community is assumed to consist of $N$ users and $M$ items (papers or other sort of scientific artifacts) which are labeled with Latin and Greek letters, respectively. The community is represented by a bipartite user-item network $\mathsf{W}$ where a weighted link between user $i$ and item $\alpha$ exists if user $i$ has interacted with item $\alpha$. Link weight $w_{i\alpha}$ is decided by the type of interaction between the corresponding user-item pair and reflects the level of importance or intensity of the interaction. It is convenient to introduce an unweighted user-item network $\mathsf{E}$ where $e_{i\alpha} = 1$ if $w_{i\alpha} > 0$ and $e_{i\alpha} = 0$ otherwise. The corresponding unweighted user and item degree are denoted as $k_i$ and $k_{\alpha}$, respectively.

We first introduce a bipartite variant of the classical HITS algorithm, biHITS, which assigns reputation values $R_i$ to user nodes and quality values $Q_{\alpha}$ to item nodes. The algorithm's definitory equations are
\begin{equation}
\label{biHITS}
\vek{R} = \mathsf{E}\vek{Q},\quad
\vek{Q} = \mathsf{E}\T\vek{R}
\end{equation}
where $\vek{R}$ and $\vek{Q}$ are \emph{user reputation} and \emph{item quality} vector, respectively. Solution to this set of equations is usually found by iterations. Starting with $R_i^{(0)} = 1/\sqrt{N}$ and $Q_{\alpha}^{(0)} = 1/\sqrt{M}$, subsequent iterations are computed as
\begin{equation}
\vek{R}^{(k+1)} = \mathsf{E}\vek{Q}^{(k)},\quad
\vek{Q}^{(k+1)} = \mathsf{E}\T\vek{R}^{(k)}
\end{equation}
and then normalized so that $\norm{\vek{R}}$ and $\norm{\vek{Q}}$ remain one. We stop the iterations when the sum of absolute changes of all vector elements in $\vek{R}$ and $\vek{Q}$ is less than $10^{-8}$. If $\mathsf{E}$ represents a connected graph, the solution is unique and independent of $R_i^{(0)}$ and $Q_{\alpha}^{(0)}$~\cite{HITS}. A weighted bipartite network can be incorporated in the algorithm by replacing the binary matrix $\mathsf{E}$ with the matrix of link weights $\mathsf{W}$.

We now simplify the QTR algorithm~\cite{Hao12} by omitting \emph{Trust} among the users---we refer it as the QR algorithm hence. Its definitory equations are
\begin{equation}
\label{QR}
\begin{split}
R_i        &= \frac1{k_i^{\theta_R}} \sum_{\alpha=1}^M w_{i\alpha}(Q_{\alpha}-\rho_Q\bar{Q}),\\
Q_{\alpha} &= \frac1{k_{\alpha}^{\theta_Q}} \sum_{i=1}^N w_{i\alpha}(R_i-\rho_R\bar{R})
\end{split}
\end{equation}
where $\bar{Q}=\sum_{\alpha=1}^M Q_\alpha/M$ and $\bar{R}=\sum_{i=1}^N R_i/N$ are the average quality and reputation value, respectively. The algorithm is further specified by the choice of $\theta_Q$, $\theta_R$, $\rho_Q$, $\rho_R$ which all lie in the range $[0,1]$. In particular, $\theta_Q$ decides whether item quality is obtained as a sum (when $\theta_Q=0$) or average (when $\theta_Q=1$) over reputation of users connected with a particular item; the meaning of $\theta_R$ is analogous. By contrast, $\rho_Q$ decides whether interactions with items of inferior quality harm user reputation (when $\rho_Q>0$) or not (when $\rho_Q=0$); the meaning of $\rho_R$ is analogous. Solution of Eqs.~(\ref{QR}) can be again found iteratively. When $\theta_Q$, $\theta_R$, $\rho_Q$, $\rho_R$ are all zero, ER differs from biHITS only in using the weighted matrix $\mathsf{W}$ instead of $\mathsf{E}$.

\subsection{Algorithms with author credit}
HITS-like algorithms that rely only on user feedback have two limitations. First, an item can only score highly after sufficient feedback has accumulated which can require substantial time in practice. Second, an item can attract the attention of users for quality-unrelated reasons (by a witty or provoking title, for example) and the algorithms lack mechanisms to correct for this. EigenRumor algorithm (ER) responds to this by introducing scores for ``information providers''~\cite{Fujimura05} which we refer to as \emph{author credit} here. While this algorithm originally includes only two sets of entities---blog entries and blog authors---it can be easily adapted to our case where users, papers, and authors are present.

The bipartite author-paper network can be represented by matrix $\mathsf{P}$ whose elements $p_{m\alpha}$ are $1$ if author $m$ has (co)authored paper $\alpha$ and $0$ otherwise ($m=1,\dots,O$ where $O$ is the number of authors). Author and paper degree in this network are $d_m$ and $d_{\alpha}$, respectively. Denoting the vector of author credit values as $\vek{A}$, the equations of EigenRumor are an extension of Eq.~(\ref{biHITS}),
\begin{equation}
\label{ER}
\vek{R} = \mathsf{E}\vek{Q},\quad
\vek{A} = \mathsf{P}\vek{Q},\quad
\vek{Q} = \omega\mathsf{P}\T\vek{A} + (1-\omega)\mathsf{E}\T\vek{R},
\end{equation}
where parameter $\omega\in[0,1]$ determines the relative contribution of authors and users to paper quality. As noted in~\cite{Fujimura05}, matrices $\mathsf{E}$ and $\mathsf{P}$ can be normalized to reduce the bias towards active users and authors. Normalization
\begin{equation}
\label{normalization}
e_{i\alpha}' = e_{i\alpha} / \sqrt{k_i},\quad
p_{m\alpha}' = p_{m\alpha} / \sqrt{d_m}
\end{equation}
is claimed to provide good results. Since the weighted user-paper interaction matrix $\mathsf{W}$ contains more information than $\mathsf{E}$, we use $\mathsf{W}'$ analogous to $\mathsf{E}'$ here.

\begin{figure}
\centering
\includegraphics[scale = 0.33]{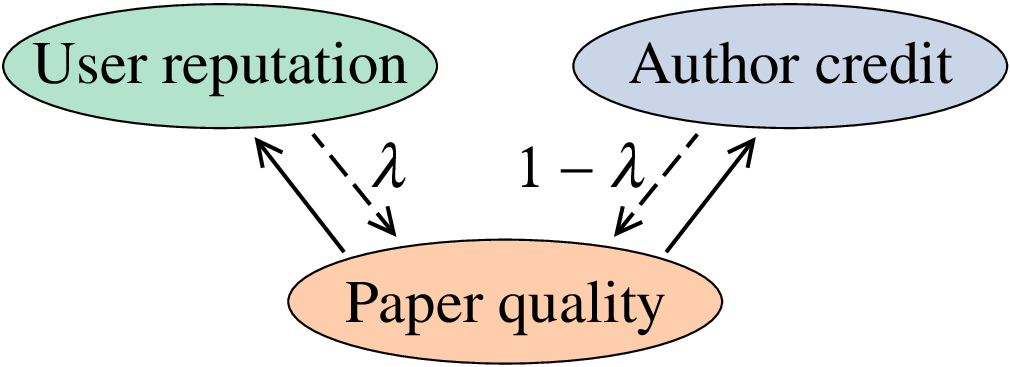}
\caption{Schematic representation of the QRC algorithm.}
\label{fig:QRCscheme}
\end{figure}

To introduce author credit in the QR algorithm and thus obtain a QRC algorithm (Quality-Reputation-Credit), we extend Eqs.~(\ref{QR}) to the form
\begin{equation}
\label{QRC}
\begin{split}
R_i        &= \frac1{k_i^{\theta_R}} \sum_{\alpha=1}^M w_{i\alpha} (Q_{\alpha}-\rho_Q\bar{Q}),\\
A_m        &= \frac1{d_m^{\phi_A}} \sum_{\alpha=1}^M P_{m\alpha} (Q_{\alpha}-\rho_A\bar{A}),\\
Q_{\alpha} &= \frac{1-\lambda}{k_{\alpha}^{\theta_Q}} \sum_{i=1}^N w_{i\alpha} (R_i-\rho_R\bar{R}) +
\frac{\lambda}{d_{\alpha}^{\phi_P}} \sum_{m=1}^O P_{m\alpha}A_m.
\end{split}
\end{equation}
Parameter $\lambda$ plays the same role as $\omega$ in EigenRumor. When $\lambda=0$, $Q_{\alpha}$ and $R_i$ are the same as obtained by QR and author credit $A_m$ is computed simply as an additional set of scores. For any other value $\lambda\in(0,1]$, all three quantities depend on each other as illustrated by Fig.~\ref{fig:QRCscheme}. Eqs.~(\ref{QRC}) can be again solved iteratively.

EigenRumor and QRC, albeit similar, show numerous differences. First, QRC uses three scores as opposed to two scores used by the original EigenRumor. Second, each summation term in QRC has its own normalization exponent ($\theta_R,\theta_Q,\phi_A,\phi_P$) which decides how to aggregate over multiple user actions, authored papers, or co-authors. The absence of explicit normalization in EigenRumor Eqs.~(\ref{ER}) is compensated by the eventual use of matrices $\mathsf{E}'$ and $\mathsf{P}'$ which makes ER's equations for $R_i$ and $A_m$ similar (up to a different value of exponent) to those of QRC. However, the ER's equation for $Q_{\alpha}$ is based on $(\mathsf{E}')\T$ and $(\mathsf{P}')\T$ which implies terms $\sum_{i=1}^n e_{i\alpha}R_i/\sqrt{k_i}$ and $\sum_{m=1}^O p_{m\alpha}A_m/\sqrt{d_m}$ without counterparts in Eq.~(\ref{QRC}).

\section{Model evaluation on artificial data}
We now describe an agent-based system~\cite{Gilbert08} which aims at producing data that can be analyzed by the benchmark QR algorithm. We aim to evaluate the algorithm's performance by comparing the true values of quality and reputation with those produced by the algorithm.

In the agent-based system, each user $i$ is endowed with intrinsic ability $a_i$ and activity level $\nu_i$, whereas each item $\alpha$ is endowed with intrinsic fitness $f_{\alpha}$. We assume that able users (those with high $a_i$) preferentially connect with high-quality items (those with high $f_{\alpha}$). Ability and activity values are both defined in $[0,1]$ and drawn from the distribution $p(x)=\mu\,x^{\mu-1}$ where $\mu\in(0,1]$ adjusts the mean value $\avg{x}=\mu/(\mu+1)$ as well as the fraction of ability/activity values above $1/2$ which is $1-2^{-\mu}$.

The system evolves in discrete time steps. At each step, user $i$ becomes active with probability $\nu_i$. In that case:
\begin{enumerate}
\item With probability $p_U$, user $i$ \emph{uploads} new item $\alpha$ to the system. The item's fitness $f_{\alpha}$ depends on the user's ability as $f_{\alpha}=a_i+(1-a_i)x$, where $x$ is a random variable drawn from $\mathcal{U}[0,X]$.
\item \emph{Downloads} two items. The probability of choosing item $\alpha$ yet uncollected by user $i$ is assumed proportional to $(f_{\alpha})^{ha_i}$ where $h>0$.
\end{enumerate}
We assume $N$ to be fixed (no new users join the community). The number of items thus grows with simulation step $t$ approximately as $M(t)=N\avg{\nu}p_Ut$ and the number of links as $E=N\avg{\nu}(2+p_U)t$. The expected network density $\eta=E/(NM)=(1+d/p_U)/N$ is thus constant.

In our simulations, we set $\mu=1/2$ so that only 30\% of users have ability/activity larger than $1/2$. We set $X=1/2$ which means that despite some level of randomness, ability of a user and fitness of items submitted by them are still related. We set $h=5$ so that users with ability close to $1$ are unlikely to accept items of low fitness (by contrast, users with zero ability accept items regardless of their fitness). Finally, we set $N=1000$ and $p_U=0.1$ which implies network density $\eta\approx 2\%$ which is similar to the values seen in real systems (while density is lower for the real data that we study here, user-item networks corresponding to the classical Movielens and Netflix datasets are of a higher density~\cite[Ch.~9]{Lu12}). We present results obtained with $t=200$ which corresponds to $\avg{M}\approx6,700$ items, $\avg{k_i}=140$, and $\avg{k_{\alpha}}=21$. Link weights assigned to uploads and downloads are $W_{\mathrm{up}}=1$ and $W_{\mathrm{down}}=0.1$ which reflects that uploading a new item is considered to be more demanding than downloading and thus deserves more reward. The influence of individual parameters on results is discussed later in this section.

To evaluate the quality and reputation estimates obtained with the algorithm, we compute the Pearson correlation coefficient between the estimated values and their true values used in the agent-based simulation: $c_{Qf}$ for items and $c_{Ra}$ for users. To assess the bias of results towards old items and active users, we measure $c_{Qt}$ and $c_{R\nu}$, respectively. While high correlation values are desirable for the first two quantities, values close to zero are preferable for the other two.

\begin{table}
\centering
\begin{tabular}{rccccc}
\hline
 Label & $(\theta_Q,\theta_R,\rho_Q,\rho_R)$ & $c_{Qf}$ & $c_{Ra}$ & $c_{Qt}$ & $c_{R\nu}$\\
\hline
biHITS &                      $(0, 0, 0, 0)$ &  $0.54$ & $0.25$ & $-0.58$ & $0.93$\\
   QR1 &                      $(0, 1, 0, 0)$ &  $0.57$ & $0.57$ & $-0.57$ & $0.15$\\
   QR2 &                      $(0, 1, 1, 0)$ &  $0.66$ & $0.61$ & $-0.46$ & $0.02$\\
\hline
\end{tabular}
\caption{Performance of three selected parameter settings in the QR algorithm.}
\label{tab:artif}
\end{table}

\subsection{Results on artificial data}
Results for the QR setting corresponding to biHITS and two other well-performing settings are shown in Tab.~\ref{tab:artif}. We see that scores obtained with biHITS correlate least with user ability and item quality and are at the same time biased towards old items and, even more, active users. BiHITS is therefore not a suitable algorithm for situations where item age and user activity are heterogeneous, which is often the case in real systems~\cite{Vazquez06,Zhou11}. While the problem of correlations between quality estimates and item age is mitigated by aging which is present in most systems of this kind~\cite{Medo11}, high correlation between user activity and reputation requires additional normalization of the biHITS algorithm as done, for example, by EigenRumor or QR.

For QR, we evaluated all 16 possible choices of parameters (two possible values---zero or one---for all four of them). The setting where $\theta_Q=0$ and $\theta_R=\rho_Q=\rho_R=1$ is the only one which is numerically unstable and thus no reportable results were obtained for it. Configurations producing the best results (see Tab.~\ref{tab:artif}) share two parameter values: $\theta_Q=0$ and $\rho_R=0$. That's not surprising as $\theta_Q=1$ would mean that popular items are not favored over unpopular ones and $\rho_R=1$ would mean that items are ``punished'' when users of low reputation connect with them. Settings QR1 and QR2 both achieve low correlation between reputation estimates and user activity which is due to $\theta_R=1$ (\emph{i.e.}, user reputation is computed as an average over user actions). The choice of $\rho_Q=1$ gives QR2 an advantage over QR1 in all four correlation metrics which means that it is indeed beneficial to punish users for uploading or downloading inferior content. The only quantity in which QR1 and QR2 perform badly is $c_{Qt}$ which is strongly negative for both but, as we already said, this is likely to be improved in real systems where aging of items results in eventual saturation of their degree growth.

We conclude with a discussion of the influence of system parameters on the results. The shape of user acceptance probability is determined by $h$. QR's performance improves with $h$ and eventually saturates at $h\simeq5$. Parameters $\mu$ and $X$ regulate the fraction of able and active users and the resulting distribution of item fitness. Our choice $\mu=0.5$ and $X=0.5$ results in able/active users being a minority and the fitness distribution being rather uniform. While $X$ is not decisive for the algorithm's performance (though, smaller values of $X$ generally lead to better results), $\mu$ is crucial as having too few able/active users makes it impossible to detect quality content. On the other hand, if able users are many, the aggregate judgment is good enough and there is no need for a sophisticated algorithm. Network sparsity $\eta$ is not particularly important as long as it is not too small (then there is too little information in the system) or too large (if every item is connected to almost all users, the presence of a link loses its information value). Finally, QR results depend only on the ratio $\xi:=W_{\mathrm{down}}/W_{\mathrm{up}}$ of the algorithm's parameters $W_{\mathrm{down}}$ and $W_{\mathrm{up}}$. When $\xi\lesssim10^{-2}$, download links are of little importance and the bipartite network effectively becomes very sparse to the detriment of the QR's performance. When $\xi\sim 1$, the performance deteriorates as well because upload information is almost neglected (note that there are many more downloads than uploads). Our original choice $\xi=0.1$ is nearly optimal.

\section{Model evaluation on real data}
Any algorithm needs to be ultimately tested by its performance on real data. To this end, we use data obtained from the Econophysics Forum (EF, see \url{www.unifr.ch/econophysics/}) which is an online platform for interdisciplinary physics researchers and finance specialists.

\subsection{Data description and analysis}
From all possible actions, we consider only interactions between users and papers uploaded to the web site: Every user can upload a paper to the site, download a paper, and view a paper's abstract. To obtain the data, we analyzed the site's weblogs created from 6th July 2010 to 31st March 2013 (1000 days in total). We removed entries created by web bots (which cause approximately 75\% of the site's traffic) and all papers uploaded before 6th July 2010 (for which we do not have the full record of user actions). To increase the data density, we removed the users who did not upload any papers and had only one action in total. In the case of a user repeatedly interacting with a given paper, only the earliest interaction was considered. Other approaches, such as cumulating all interactions or preferring paper downloads over abstract views, for example, result in inferior performance of QR. This choice is further motivated by the fact that the first interaction does best represent the user's interest: Papers that really capture users' attention are downloaded/read immediately when encountered, whereas a later download indicates other reasons of interest. The final input data contains 5071 users, 844 papers and 29748 links, implying $\eta\approx 0.7\%$. Note that the Econophysics Forum has an editor who has uploaded 85\% of all papers in the analyzed sample. Paper metadata includes paper submission time, title, and a list of its authors. To avoid the problem of an author's name represented in multiple ways (\emph{e.g.}, 'H.~Eugene Stanley' vs 'H.~Stanley' vs 'HE~Stanley'), we use only the first initial without comma and the surname ('H Stanley'). As a result, there are 1527 authors in the analyzed sample. The paper metadata was augmented by citation counts, which were obtained from Google Scholar on 12th December 2013, and by impact factors of the journals where papers were eventually published. We shall use this external information to evaluate rankings of papers produced by various algorithms.

\begin{figure}
\centering
\includegraphics[scale = 0.3]{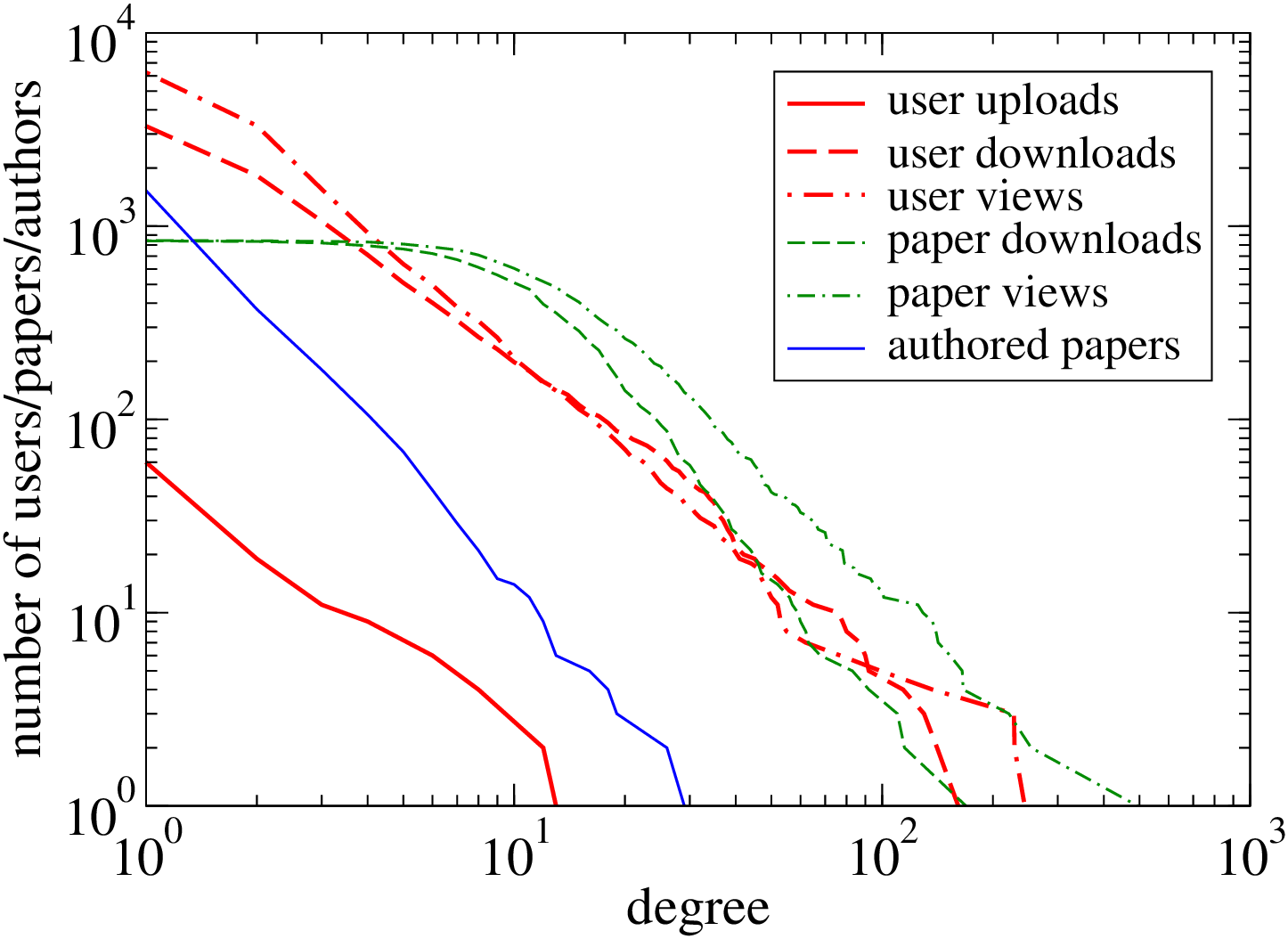}
\caption{Cumulative degree distributions in the Econophysics Forum data with respect to various actions for users, papers, and authors. The editor was removed from the user upload distribution for the sake of clarity.}
\label{fig:deg_distrib}
\end{figure}

Figure~\ref{fig:deg_distrib} shows cumulative degree distributions for all involved parties: Users, papers, and authors. All distributions are broad and some of them might even pass statistical tests for power-law distributions. As a result, while 92\% of users have ten actions in total or less, the most active users downloaded or viewed roughly a hundred of papers. With respect to the time span of the data, this is still a human level of activity which suggests that our removal of automated access was reasonably successful. The degree distribution of papers is shifted to the right as a whole with a negligible number of papers downloaded or viewed less than ten times and the most successful papers being of interest to hundreds of users. The most active authors are well-recognized in the econophysics community: Jean-Philippe Bouchaud, Shlomo Havlin, Dirk Helbing, Didier Sornette, and Eugene Stanley (in alphabetical order) have all authored more than 15 papers in the sample.

\subsection{Results on real data}
To distinguish the three different actions (upload, download, and abstract view), we set the respective link weights $W_{\mathrm{up}}=1$, $W_{\mathrm{down}}=0.1$, and $W_{\mathrm{view}}=0.05$. This acknowledges paper upload as the most demanding activity and viewing an abstract signalizes paper quality less than its direct download.

\begin{table}
\centering
\begin{tabular}{rr@{$\,\pm\,$}lr@{$\,\pm\,$}lr@{$\,\pm\,$}lr@{$\,\pm\,$}l}
\hline
 Label & \multicolumn{2}{c}{Day} & \multicolumn{2}{c}{Down} &
         \multicolumn{2}{c}{Cit} & \multicolumn{2}{c}{IF}\\
\hline
  RAND & $548$ & $41$ & $11$ & $1$  &  $5$ & $1$  & $1.1$ & $0.2$\\
   POP & $299$ & $37$ & $69$ & $7$  & $15$ & $4$  & $1.5$ & $0.5$\\
biHITS & $264$ & $34$ & $56$ & $7$  & $10$ & $3$  & $1.4$ & $0.3$\\
    ER & $444$ & $49$ & $30$ & $10$ & $18$ & $4$  & $2.7$ & $0.5$\\
   QR1 & $375$ & $49$ & $59$ & $9$  & $15$ & $4$  & $1.9$ & $0.6$\\
   QR2 & $445$ & $47$ & $54$ & $9$  & $14$ & $3$  & $1.9$ & $0.6$\\
   QRC & $465$ & $60$ & $34$ & $8$  & $34$ & $10$ & $3.8$ & $0.6$\\
\hline
\end{tabular}
\caption{Mean and standard error for basic metrics (submission day, number of downloads, citation count, and journal impact factor) of top 20 papers obtained with various algorithms (ER uses $\omega=0.20$, QRC uses $\lambda=0.57$).}
\label{tab:real}
\end{table}

We begin our analysis by inspecting algorithms without author credit: Random ranking of papers (RAND), popularity ranking (POP), where popularity is measured by the number of downloads, and biHITS. The average characteristics of top twenty papers according to these and other methods are summarized in Tab.~\ref{tab:real}. The expected bias towards old papers is clearly visible for the POP ranking whose top papers are on average 8 months older than RAND papers. While mean citation count of popular papers exceeds that of random papers, two of the most popular papers have never been published and four have not been cited to date: Wisdom of the crowd appears to be no good guide here. Both RAND and POP provide no information on the ranking of authors. BiHITS shows stronger bias towards old papers than POP which is not surprising as it is, ultimately, also a popularity-driven algorithm. Furthermore, it awards the Econophysics Forum editor who uploaded majority of papers with score which is so high that views and downloads by ordinary users add only small variations to the score of those papers. Even worse, papers that have not been submitted by the editor cannot reach the top of the ranking regardless of their success among the users. Thanks to normalization, the editor's weight does not represent a problem in QR1 and QR2. On the other hand, their top papers are not cited more than papers chosen by biHITS or POP. Furthermore, QR1 and QR2 choose rather popular papers and one could argue that they actually provide little new and useful information to the users. In fact, the excessive tendency of information-filtering algorithms towards popular objects is a long-standing challenge in this field~\cite{Ado12,Gualdi13}.

Before analyzing ER and QRC, the parameters of QRC need to be set. We use $\theta_Q, \theta_R, \rho_Q, \rho_R$ corresponding to QR1 which performed best on artificial data. We have also evaluated a variant of QRC based on QR2 and found that penalization of users connected to low quality papers through $\rho_Q=1$ leads to negative paper scores and in turn various counter-intuitive results. To avoid assigning high credit to authors of a single successful paper (beware the trap of papers with attractive titles), we use $\phi_A=0$ which results in accumulation of author credit over the course of time. Since $\phi_P=0$ (summing the credit of a paper's authors) gives an advantage to papers with many authors, we use $\phi_P=1$. We have evaluated other possible choices of parameters $\phi_P,\phi_A$ (as well as some other choices, such as paper quality contributed by the sum of credit of two most credible authors) and found that $\phi_A=0$ and $\phi_P=1$ indeed produce the most satisfactory results.

\begin{figure}
\centering
\includegraphics[scale=0.3]{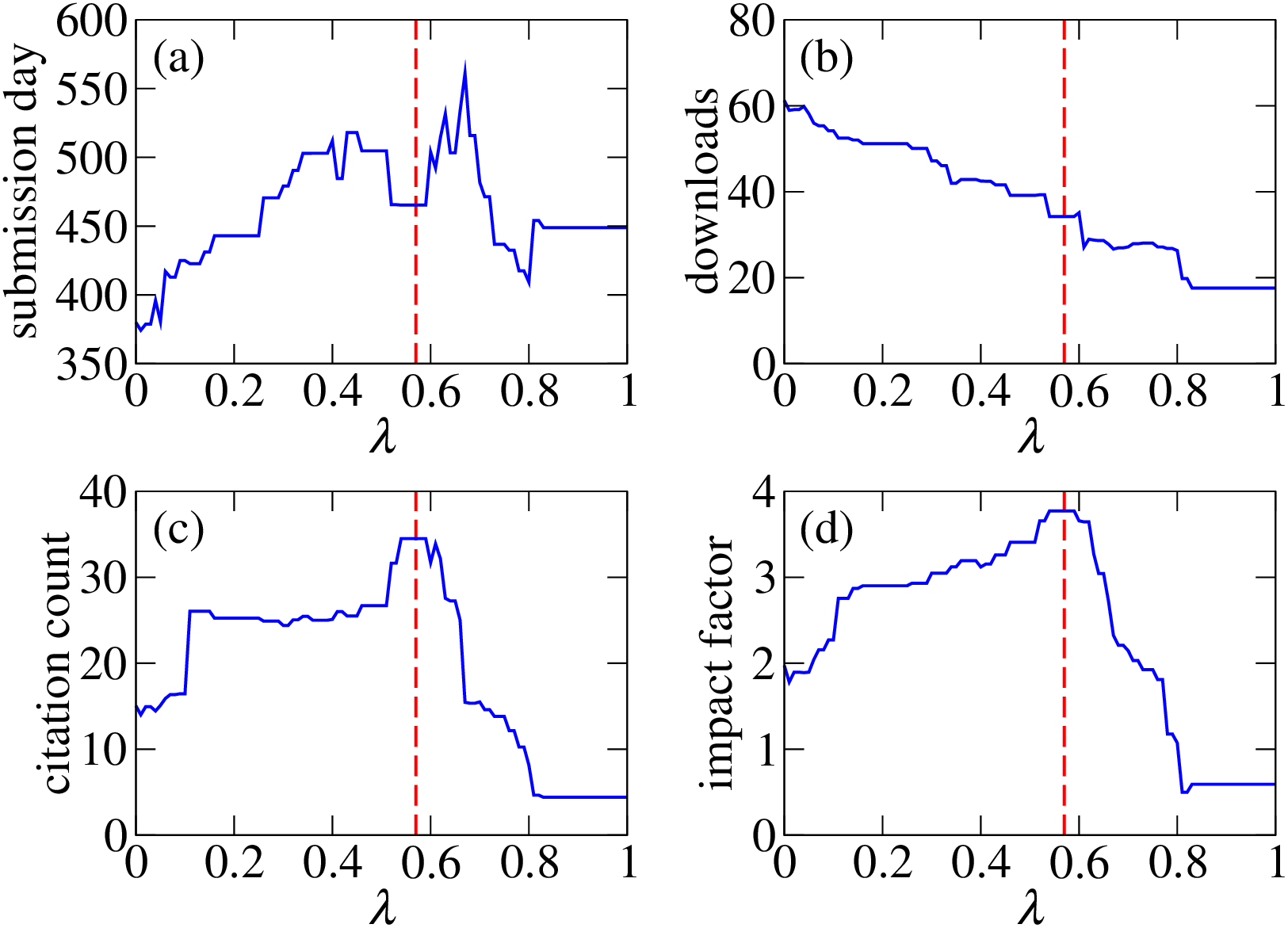}
\caption{Average metrics of QRC's top $20$ papers versus $\lambda$. The vertical dashed line at $\lambda=0.57$ marks the setting where citation count and impact factor are approximately maximized.}
\label{fig:lambda}
\end{figure}

Fig.~\ref{fig:lambda} shows the average metrics of the top twenty papers obtained with QRC for $\lambda\in[0,1]$. As $\lambda$ increases, the average submission day of papers in top $20$ grows from $375$ (the original QR1 value) to $519$ when $\lambda=0$; the inclusion of author credit thus helps to mitigate or even remove the time bias. The average number of downloads decreases with $\lambda$ and eventually reaches less than 25\% of the QR1 value. The average impact factor is improved over a wide range of $\lambda$ and peaks at $3.8$ for $\lambda\approx0.57$. The same is true for the average citation count which peaks at $34$ for $\lambda=0.57$. As can be seen in Table~\ref{tab:real}, QRC outperforms the other evaluated methods. The Mann-Whitney U test based on top 20 papers chosen by various algorithms confirms that QRC outperforms them at the significance level $0.02$ with the exception of ER where, due to the small sample size and large fluctuations, significance is only $0.08$.

There are two further points to make. First, top papers chosen by QRC are generally younger than those chosen by other methods and thus have had less time to accumulate citations. Second, QRC is the only method which puts ``Catastrophic Cascade of Failures in Interdependent Networks'' (available on arXiv under ID 1012.0206) among the top papers. This paper with mere three citations is a summer-school version of a slightly earlier identically entitled work which has accumulated almost 500 citations (it has not been submitted to the Econophysics Forum). The paper's small contribution to the overall citation count achieved by QRC thus severely underestimates the paper's true importance. In summary, QRC's overall citation count improvement is thus likely to be underestimated.

Since citation counts alone provide imperfect information about the quality of scientific work, we now turn to authors. Table~\ref{tab:authors} lists top ten authors obtained by QRC with $\lambda=0.57$ to show that they indeed include reputed names from the field of econophysics (another outstanding person, J.-P. Bouchaud, is just below the line, on place 11, largely because of collaborating with less active and thus less QRC-credible co-authors). As of December 2013, the mean $h$-index of QRC's top 10 authors obtained by querying the Thomson's Web of Knowledge was $41\pm11$ which is significantly more than $4\pm2$ for top 12 authors (who all have identical credit) according to EigenRumor.

\begin{table}
\centering
\begin{tabular}{rlrrr}
\hline
Rank & Name           & Credit & Papers & Down\\
\hline
   1 & H. E. Stanley  & $0.65$ &     26 & 22\\
   2 & T. Preis       & $0.39$ &      8 & 38\\
   3 & D. Sornette    & $0.35$ &     29 & 17\\
   4 & S. Havlin      & $0.22$ &     19 & 11\\
   5 & B. Podobnik    & $0.19$ &      8 & 21\\
   6 & D. Y. Kenett   & $0.16$ &     11 & 14\\
   7 & D. Helbing     & $0.16$ &     18 & 20\\
   8 & E. Ben-Jacob   & $0.14$ &     10 & 12\\
   9 & A. M. Petersen & $0.10$ &      6 & 13\\
  10 & S. V. Buldyrev & $0.09$ &      7 & 13\\
\hline
\end{tabular}
\caption{Top ten authors in the QRC ranking: Their credit, number of authored papers, and the average number of downloads. The overall average number of papers per author and downloads per paper are $1.6$ and $13$, respectively.}
\label{tab:authors}
\end{table}

\section{Discussion}
We have proposed QRC, a new reputation algorithm for scientific online communities. QRC is based on three main components: \emph{Quality} of papers, \emph{Reputation} of users, and \emph{Credit} of authors. We have used data from a scientific community web site, the Econophysics Forum, to evaluate the algorithm and compare its performance with that of other reputation algorithms. The newly proposed QRC algorithm outperforms those algorithms in various aspects. Papers scoring high in the resulting QRC algorithm are younger than those selected by bipartite HITS and they have been downloaded considerably fewer times than papers selected by any other algorithm considered here. At the same time, QRC's top papers have attracted significantly more citations and the average impact factors of their publication venues is also higher than for papers chosen by the other algorithms. In short, QRC is able to highlight the papers that have been largely neglected by the Econophysics Forum users (as demonstrated by their relatively low number of downloads), yet they have eventually attracted considerable attention from the scientific community (as indicated by the publication venues and the citation counts). Note that QRC introduces author credit endogenously, relying on no other information than user activity on the given web site. The observed improvements are thus not achieved by providing this algorithm with more information than what is made available to the other algorithms. Finally, QRC's top authors have on average substantially higher $h$-index than top authors found with other algorithms.

In the context of predicting future citation counts of papers, QRC represents an algorithm-focused alternative to machine-learning approaches~\cite{Castillo07,Acuna12}. Note that the algorithm's range of applicability is not strictly limited to scientific online communities. QRC can be used in any community where: (1) shared perceptions of quality can emerge, (2) quality induces popularity, and (3) individual items have various authors. If a scientific community is in divide, for example, and its members deeply disagree on some theories or methods, condition (1) is violated and an attempt to produce a universal quality ranking might be in vain. While the causality between quality and popularity in science is imperfect (effects such as the first-mover advantage have reported~\cite{Newman09}), it is still stronger than in music, for example, where condition (2) is questionable and the use of QRC is likely to produce dubious results. To overcome these limitations and thus extend the QRC's range of applicability remains a future challenge.

There are several research directions which remain open. The behavior and performance of the QRC algorithm upon non-integer choices of its parameters (such as the exponent $0.5$ used in~(\ref{normalization})) need to be examined. User surveys can be employed as an additional evaluation tool complementing the current quantitative approach based on citations, impact factor and $h$-index. As any other reputation system, QRC has also inherent preferences (and thus also incentives) for various kinds of behavior. In particular, it favors active authors and those who collaborate with other credible authors. Various forms of gaming of research metrics, ranging from self-citations and mutual citations to plagiarism therefore deserve particular attention. Thanks to its reliance on long-term quality indicators (author credit), QRC has the potential to prove substantially more robust to malicious behavior than its predecessors. For input data exceeding the three-year time span of the presently studied Econophysics Forum data, it may be suitable to introduce time decay of quality and credit values to prevent the oldest contributions and the most active authors from occupying top positions in their respective rankings. Results presented in~\cite{Medo11,Wang13} may provide a starting ground for these efforts. One should not forget that the QRC results are community-specific as they are based on feedback of a given group of users. This is not only a limitation but also an opportunity: The QRC algorithm can be eventually used to study the dynamics and differences between various research communities.

\begin{acknowledgments}
This work was supported by the EU FET-Open Grant No.~231200 (project QLectives) and by the Swiss National
Science Foundation Grant No.~200020-143272.
\end{acknowledgments}


\begin{thebibliography}{99}
\bibitem{Radicchi08} F. Radicchi, S. Fortunato, C. Castellano, Proc. Nat. Acad. Sci. 105, 17268 (2008)
\bibitem{Lari08} V. Larivière, É. Archambault, Y. Gingras, J. Am. Soc. Inform. Sci. Tech. 59, 288 (2008)
\bibitem{Maxmen10} A. Maxmen, Cell 141, 387 (2010)
\bibitem{Resnick00} P. Resnick, R. Zeckhauser, E. Friedman, K. Kuwabara, Commun. ACM 43, 45 (2000)
\bibitem{Sabater05} J. Sabater, C. Sierra, Artif. Intell. Rev. 24, 33 (2005)
\bibitem{Axelrod} R. Axelrod, The Evolution of Cooperation (Basic Books, 1984)
\bibitem{Massum04} H. Masum, Y.-C. Zhang, First Monday 9, no. 7 (2004)
\bibitem{Josang07} A. J\o sang, R. Ismail, C. Boyd, Decis. Support Syst. 43, 618 (2007)
\bibitem{Hanrahan12} B. V. Hanrahan, G. Convertino, L. Nelson, Proc. ACM 2012 Conf. Comp. Supp. Coop. Work Companion, 91 (ACM, 2012)
\bibitem{Newman_book} M. Newman, Networks: an introduction (Oxford University Press, 2009)
\bibitem{PR} S. Brin, L. Page, Comput Networks ISDN 30, 107 (1998)
\bibitem{Franc11} M. Franceschet, Commun. ACM 54, 92 (2011)
\bibitem{HITS} J. Kleinberg, J. ACM 46, 604 (1999)
\bibitem{Fujimura05} K. Fujimura, N. Tanimoto,
In R. Falcone et al. (Eds.): Trusting Agents for Trusting Electronic Societies, 59 (Springer-Verlag, 2005)
\bibitem{coHITS} H. Deng, M. R. Lyu, I. King, Proc. ACM 15th Int. Conf. Knowl. Disc. Data Mining, 239 (ACM, 2009)
\bibitem{Hao12} H. Liao, G. Cimini, M. Medo, In L. Chen et al. (Eds.): ISMIS 2012/Lect. Notes Artif. Int. 7661, 421 (Springer-Verlag, 2012)
\bibitem{Gilbert08} N. Gilbert, Agent-based Models (SAGE Publications, 2008)
\bibitem{Lu12} L. Lü, M. Medo, C. H. Yeung, Y.-C. Zhang, Z.-K. Zhang, T. Zhou,
Phys. Rep. 519, 1 (2012)
\bibitem{Vazquez06} A. Vázquez, J. G. Oliveira, Z. Dezsö, K. I. Goh, I. Kondor, A. L. Barabási,
Phys. Rev. E \textbf{73}, 036127 (2006)
\bibitem{Zhou11} T. Zhou, M. Medo, G. Cimini, Z.-K. Zhang, Y.-C. Zhang,
PLoS ONE 6, e20648 (2011)
\bibitem{Medo11} M. Medo, G. Cimini, S. Gualdi,
Phys. Rev. Lett. 107, 238701 (2011)
\bibitem{Ado12} G. Adomavicius, Y. Kwon,
IEEE Trans. Knowl. Data Eng. 24, 896 (2012)
\bibitem{Gualdi13} S. Gualdi, M. Medo, Y.-C. Zhang, EPL 101, 20008 (2013)
\bibitem{Castillo07} C. Castillo, D. Donato, A. Gionis,
In N. Ziviani, R. Baeza-Yates (Eds.): SPIRE 2007/Lect. Notes Artif. Int. 4726, 107 (Springer-Verlag, 2007)
\bibitem{Acuna12} D. E. Acuna, S. Allesina, K. P. Kording, Nature 489, 201 (2012)
\bibitem{Newman09} M. E. J. Newman, EPL 86, 68001 (2009)
\bibitem{Wang13} D. Wang, C. Song, A. L. Barabási, Science 342, 127 (2013)
\end{thebibliography}
\end{document}